# Photonic Origami of Silica on a Silicon Chip with Microresonators and Concave Mirrors

MANYA MALHOTRA, RONEN BEN-DANIEL, FAN CHENG, AND TAL CARMON*

*School of Electrical Engineering, Tel Aviv University, Tel Aviv 6997801, Israel*
*total@tauex.tau.ac.il

**3D printing of high-quality silica photonic structures is particularly challenging, as surface roughness at the nanoscale can severely degrade optical performance through scattering losses. Here, we develop a technique to fold ultrasmooth silica on silicon chips into such desired 3D structures. A laser-induced, surface-tension-driven method achieves folding with 20 nm alignment accuracy, enabling origami-like polylines and helices with integrated 0.5 nm-smooth photonic devices. The technique allows for the fabrication of record length-to-thickness ratio structures, incorporating concave micromirrors with numerical aperture of 0.41, and microresonators with quality factors exceeding $8\times10^6$. This on-chip silica origami approach offers a pathway to transform planar electro-opto-mechanical circuits into high-quality 3D configurations.**

**Introduction:** silica is essential in science and society and is renowned for its superior properties. In 1966, Kao extrapolated the absorption Lorentzians profiles of several glasses toward their tails, which led to the prediction that silica would have ultrahigh transparency if purified [1]. This discovery enabled fibers to become the backbone of modern communication. Going from 1D fibers to planar 2D photonics involved replacing the fiber's solid cladding with air, which resulted in a higher core-clad index contrast. This index contrast increased the scattering of light at the solid-phase boundaries of resonators and therefore had to be addressed. Accordingly, reflow and etching processes were developed to control silica smoothness down to 0.15 nm [2–4], enabling ultrahigh Q [UHQ] photonics [2,3,5–13] with Q exceeding $8\times10^9$ [2]. Such advancements supported applications including micro gyroscopes [5,6], atomic clocks [7,8], frequency combs [9,10], lidar [11,12], and label-free single-molecule detectors [13]. These micro-devices [5–13], as well as our pre-folded structures, are fabricated in cleanrooms and fabs in sequential processing steps and with 0.5 nm smoothness [4]. Such steps, which can overlap during execution, make the fabrication workflow both efficient and scalable.

The next natural step was the transition from 1-2D configurations to 3D ones, which involved going from lithography, a subtractive fabrication process that is, again, smooth at the ultimate atomic scale [4], to 3D printing—an additive process relying on building objects layer by layer. Such 3D printing was demonstrated in glasses [14–18]. Nevertheless, 3D printing of smooth, uniform, isotropic, and clean devices from glass is challenging due to the inherent voxels that are added in a discrete manner while relying on photocurability [18], heating [15], or nonlinear laser absorption [19] processes. Such photo-fabricated voxels typically solidify in proportion to the optical intensity squared, inherently resulting in rough surfaces that scatter light. As a possible alternative to printing, technologies have emerged for folding 2D planes into 3D structures; among them are folding techniques relying on elasto-capillarity [20], residual stress [21,22], mechanical stress [23,24], responsive forces [25,26], and focused-ion-beam methods [27,28]. These techniques have been applied to large metals [29], glasses [30], and silicon structures [31]. Origaming of 2D lithographed silica on a silicon chip, as we demonstrate in what follows, might enable 3D circuits with ultra-high Qs, but it has rarely been studied.

Here, we demonstrate a fast process for folding silica sheets on a silicon chip in less than a millisecond and with acceleration exceeding 19,600 m/s$^2$. We do so by borrowing hygroscopic principles used in plant, including in seed-release mechanisms to fold photo-lithographed silica by harnessing surface tension. Using a laser, we selectively liquefy [3] one side of a silica sheet to fold it in a touchless, additive-free manner. Furthermore, by avoiding wet etching while preparing the bars to be folded, we achieve the ultimate length-to-thickness ratio, [Fig. 1(a)]. Despite some shape restrictions, our silica origami on silicon chips might impact opto-electro-mechanical circuits by combining nano-alignment, record length-to-thickness ratios, and UHQs. Long-term visions include origami as a standard cleanroom technology for shaped light sails [32], folding self-healing glass [33], origami-enabled smart materials [34], developable patterns [35], micro-robotic arms [36] as well as integrated 3D circuits with reflowed angstrom-smooth concave [13], toroidal [3], and spherical [2] optics.

**Experimental setup:** we start with an amorphous silica layer on a silicon chip. Specifically, silica is thermally grown on a silicon chip using cleanroom-grade silicon and oxygen. As for the silica surface smoothness, our fabrication process [4] produces silica surfaces with smoothness down to 0.5 nm, as described in reference [4]. Lastly, before bending, we use dry Xenon Difluoride [$XeF_2$] silicon etching to release the silica from the silicon while leaving a supporting region to hold the silica. This process is generally referred to as undercutting the silica. It is crucial that we use a dry $XeF_2$ etching process because if we had used wet etching, interfacial tension at the liquid-solid phase border would have broken the long and thin silica sheets. In contrast, by using dry etching, our cantilever silica bars can be ultra-thin yet very long, (See Supplement 1).

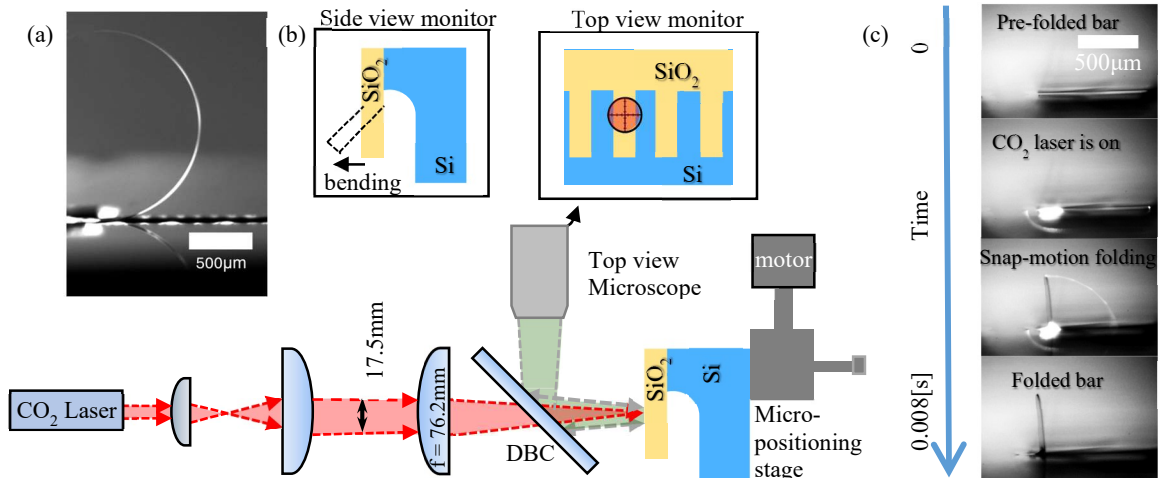

**Fig. 1 Experimental setup (a) Ultrahigh slenderness ratio** of a 3 mm long and 500 nm thick silica bar **(b) Experimental setup** for laser-induced bending of silica on a chip. Light from a $CO_2$ laser focuses on the desired folding region. A crosshair (Red) on a top-view microscope monitor marks the place where the laser will meet silica. A micropositioning stage is used to choose the folding region. In addition, a side-view microscope monitors the bending process. DBC: Dichroic Beam Combiner. **(c) Snap-motion folding:** The photographs represent a horizontal 1 mm long, 5 µm thick silica bar on a silicon chip rapidly bending to vertical right after the laser is turned on. The laser pulse is 4 ms long and at 26.5 KW/cm² intensity. The molten silica turns into a radiating blackbody, as seen in the bright region. In the third photo, light from the molten region is guided through the bar and scatters from its tip, creating an arc-shaped flare. See a slow-motion movie describing the snap-motion folding process in Supplementary Video 1. The camera frame rate is 500 frames per second.

The slenderness ratio, s, of a cantilever beam relates to the ratio of its length, L, to its thickness, h, which gives s = 6,000 for our L = 3 mm, h = 0.5 µm bar [Fig.1(a)]. This is in comparison to a slenderness ratio of 3000 reported for a single-crystal diamond structure [37]. To give a sense of the upper limit on the slenderness ratio, doubling the length of our beam would result in a 2.3 mm deflection under gravitational forces, which is 40% of its length (See supplement 2). This deflection, under gravitational attraction only, indicates the upper limit on the bar's slenderness ratio. Such a large slenderness ratio permits structures having an ultimate size-to-weight ratio that is helpful for light sails [32] and for testing gravity models [38,39] using optically levitating cavity mirrors [40]. In this regard, we fabricate 3 mm long structures that are only 0.5 µm thick [Fig. 1(a)].

Generally speaking, we find our folding process to be highly versatile across a broad range of silica thicknesses. That said, some thickness considerations should be taken into account. Specifically, when the thickness is below 0.5 µm, the sample naturally bends due to thermal stress related to the process of silica fabrication via thermal oxidation of silicon. This bending can be mitigated by relying on the sol-gel silica growth process [41], as an alternative. As the sample becomes thicker, it suffers less from thermal stress. We could fold silica sheets up to 10 µm in thickness, limited by wafer availability, as well as 125 µm diameter fibers. Our experience suggests that bars thicker than 125 µm could also be easily folded; yet, in such thick samples, one might want to consider gravity and align it with the direction of surface tension forces.

Fig. 1(b) schematically shows our setup used to fold silica sheets on a silicon chip in a controllable manner. We use an 11 µm wavelength laser to selectively melt the upper side of a silica microbar, bringing it to nearly 3000 K temperature while using microscopes to control the process. Such bending occurs in less than 1 ms [Fig. 1(c)]. In this process, we harness interfacial tension at the liquid-gas phase border to bend the bar against gravity and against some viscosity. Viscosity originates mainly from the cooler side of the bar, where the temperature is only slightly above the glass transition temperature (1500 K). When we need to bend to a specific angle at high resolution, we send a train of relatively low-power pulses and stop at the required angle. A slow train of pulses permits conventionally stopping at the required angle. Additionally, we can make several knees on a single bar, turning it into a polygon. Furthermore, continuous folding into a circular or helical shape is achieved by monotonically moving the sample with respect to the laser focus. Since similar viscosity and surface tension characteristics at liquid-phase boundaries above glass transition temperature can be observed in many materials, we believe that other materials can also be folded, as exemplified here for silica. For example, we believe that dielectrics such as silicon nitride and aluminum oxide, as well as compound semiconductors such as amorphous gallium arsenide (a-GaAs), are capable of being folded using a laser.

In detail, our experimental setup [Fig. 1(b)] consists of a $CO_2$ laser beam, focused on a 60 µm full width at half maximum [FWHM] spot at the region we intend to fold. We use a top-view microscope, bore-cited (aligned) with the laser, to bring the silica region that we intend to bend into the optical focus of the laser. Simultaneously, a side-view microscope allows for the monitoring and control the bending angles.
**Results:** The photographs in Fig. 1(c) and 2(a), along with the movie (Supplementary Video 1), depict a typical bending event where laser pulses initiate a snap-motion folding of 90°. This occurs at an acceleration of 2004 g in less than a millisecond while reaching a speed of 2 m/s. Such rates are fast compared to other micro folding processes [42]. A short time constant benefits the fast duration of the folding process. We estimate the thermal time constant of the molten region to be shorter than 30 µs, based on the ratio of heat capacity to total thermal conductance (including contributions from conduction, convection, and radiation). The time constant relates to the time it takes for the silica to solidify once the $CO_2$ laser is turned off. A closer look at the knee region [Fig. 2(a) inset] reveals a frozen droplet shape, typical of situations where interfacial tension at the liquid-phase boundary minimizes surface energy by reducing interface area. In this context of rapid motion, we calculate that the energy gained from interfacial tension is sufficient to lift a typical folded bar, against gravity, to a height greater than 1 meter. Therefore, this surface tension energy is much larger than the sub-millimeter gravitational potential energy that the bar gains by bending. In summary, the bending process is rapid because surface tension energy far exceeds gravitational potential energy, and it ends quickly thanks to the short thermal time constant.
**Folding to an arbitrary angle:** to fold to an arbitrary angle, we reduce the $CO_2$ laser power and divide it into a train of pulses [Fig. 2(b) and movie in Supplementary video 2] so that we can stop when reaching the desired angle.
**Fine-tuning the bending angle:** for fine-tuning the bending angle [Fig. 2(c)], we reduce the intensity until the bending per pulse is minimal. We have achieved a bending resolution of 0.1 milliradians per pulse, corresponding to nanometers position-control resolution for a typical arm length.
**Discrete foldings into a polyline and continuous folding into a helix shape:** as one can see in Fig. 2(d)- 2(e), we fold the bar at multiple points, either as a series of discrete bends [Fig. 2(d)] or through continuous folding into a helical shape [Fig. 2(e)] by monotonically moving the bar while heating the bending region. Notably, the last bend in Fig. 2(d), at the lowest point in the image, is performed while the bar obscures a large portion of the laser beam. Despite this obstruction, bending remains possible by increasing the laser power.

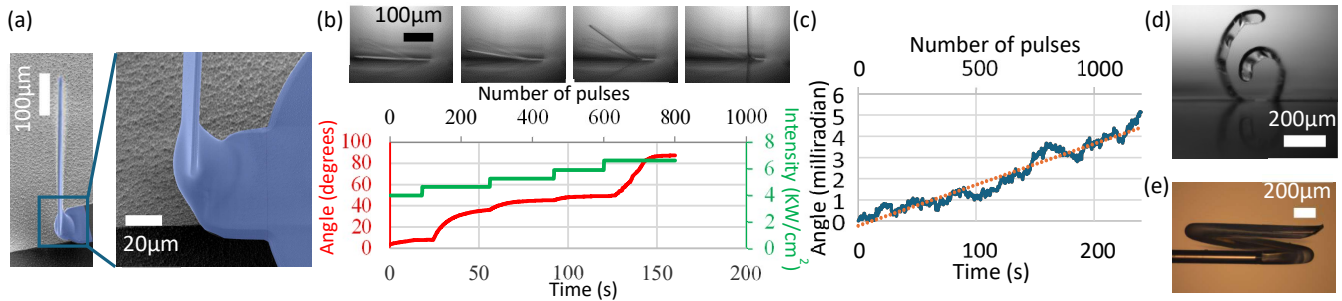

**Fig. 2 Experimental results in bending silica.** (**a**) **90° bending** within less than 1 ms by applying 26.5 KW/cm² intensity for 4 ms (**b**) **Continuous folding** (Red) by applying a train of 2 ms pulses. at increased intensity (Green) to compensate for the fact that the bending bar starts hiding the light. (**c**) **Fine-resolution bending** using a train of pulses, each of 2.55 KW/cm² intesity and 1.5 ms duration. The fluctuation in of the experimental points are due to the limitation of the tracking software. Our bending resolution is estimated at 0.1 milliradian. (**d**) **Bending in discrete steps**. 50 pulses of 1 ms for each knee (see Supplement 3 for further details). (**e**) **Helical bending** is achieved by moving the bent structure at a 0.55 mm/s speed through an intensity of 11.9 KW/cm².

**Parabolic mirrors: Incorporating concave nano-smooth surfaces into the origami via evaporative evacuation of material:** micro-parabolic mirrors [43,44,13] are a significant building block in photonics and have been demonstrated to enable Fabry-Pérot resonators with record finesse values exceeding $10^6$ [43]. Similar to previous works [43,44,13], we fabricate a concave parabolic surface on silica origami by using a CO2 laser to reshape the flat surface into a paraboloid through controlled evaporation of the molten glass [Fig. 3(a)], revealing a focal length of 38 µm at a 34 µm diameter corresponding to a numerical aperture [NA] of 0.41. Such paraboloids have achieved the highest Fabry-Perot finesse [43].
**Dielectric whispering-gallery resonators: Incorporating concave (spherical) nano-smooth surfaces into the origami via material reflow:** here, we incorporate a dielectric sphere into the origami [Fig. 3(b)] and activate it as a fiber-coupled whispering-gallery resonator. Material removal via evaporation is suitable for concave surfaces, where the structure increases its surface area during formation. In contrast, fabricating convex shapes uses surface tension to reduce surface area as they form. Accordingly, we incorporate convex devices into the origami by relying on surface tension. In detail, we reflow a folded silica bar, on-chip, using a $CO_2$ laser to form a dielectric sphere. The spherical resonator's

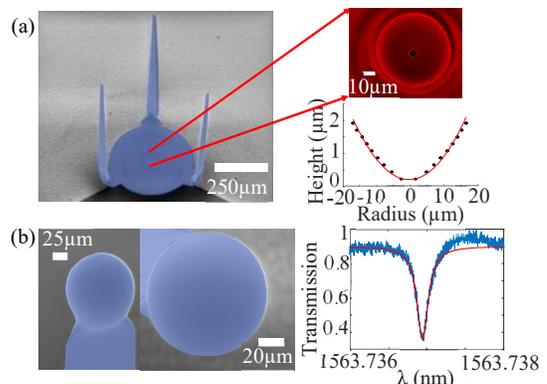

**Fig. 3 Incorporating convex and concave devices into the origami** (**a**) A table with a parabolic reflector incorporated. The interferogram reveals the high map (blue circles) with parabolic best fitting (red line) revealing a 38 µm focal distance (**b**) A spherical resonator with transmission indicating an optical quality factor of 8.7262 x $10^6$ at 1550 nm wavelength.

quality factor tested using a tapered fiber coupler [45] was $8.73\times10^6$, as anticipated from the resonance spectral width.

Using two complementary techniques, removal of material via evaporation and reflow to accumulate material, we demonstrate [Fig. 3] the incorporation of both concave and convex nano-smooth photonic devices in our silica origami (parabolic micromirrors and dielectric whispering-gallery resonators). To give just one example, the table structure in Fig. 3(a), which will be fully released later, has a parabolic surface at its center and can serve as a levitating cavity mirror, similar to what was theoretically suggested [40] for optical levitation of cavity mirrors [40] for testing gravity models [38,39]. In this regard of using optical tweezers to levitate macroscopic objects, the table's small weight-to-size ratio and its nano-smooth concave paraboloid could benefit such proposed cavity-mirror levitators [40]. Photonic origami also enables the fabrication of convex and concave microlens arrays and monolithic on-chip cascaded whispering-gallery-mode (WGM) resonators.

**Discussion:** while most 3D manufacturing techniques rely on additive fabrication in discrete steps, leading to challenges in uniformity, purity, and smoothness, we combine subtractive 2D lithography with tunable bending, by harnessing surface tension, to fabricate 3D structures that are ultrasmooth. We demonstrated bending from 0° to unlimited helices [Fig. 2(e)] with 0.1 milliradians resolution [Fig. 2(a)- 2(c)] and 20 µm minimal curvature radius. The upper radius limit relates to our ability to produce 0.1 milliradian knees at a specific location.

The surface energy gained during the bending process exceeds the gravitational potential energy required to bend a typical bar by more than 2,000 times, while the thermal time constants are under 30 µs, facilitating rapid bending. Furthermore, we demonstrate that our 3D structures are compatible with reflow and evaporation techniques, enabling the addition of concave and convex ultra-smooth surfaces used in Fabry-Perot and whispering gallery resonators. The highest optical transmission of silica, combined with its compatibility with on-chip electronics, along with photonic reflow and removal of material via evaporation make our origami, highly versatile. Additionally, the scalability of lithography ensures adaptability to specific needs in both applied and fundamental studies, that our origami technique extends here to 3D circuits.

It seems about time now, as silica quality has reached the fundamental limitations of smoothness and purity, and with our origami enabling the breakthrough of the planar 2D-barrier, that 3D ultrahigh Q circuits will become everyday tools of science. Origaming lithographed silica on silicon with smoothness and nano-precision alignment might hence permit integrated 3D electro-opto-mechanical circuits with ultrahigh quality factors, while being compatible with scalable manufacturing, and fabrication flexibility.

**Funding.** United States–Israel Binational Science Foundation (NSF-BSF) (2020683); Israel Science Foundation (537/20).

**Acknowledgment.** We acknowledge help from Sebastian Hofferberth, Lucas Tenbrake, Stefan Linden, Hannes Pfeifer, Bar Reuven, Zahava Barkay, Amit Shacham, Nicole Gorokhovsky, and Anastasia Adelberg.

**Disclosures.** The authors declare no conflicts of interest.

**Data availability.** All data, code and materials used here are available to any researcher upon request.

**Supplemental Video** 1: See Supplement 1 for supporting content.

**References**
1. K. C. Kao and G. A. Hockham, Proceedings of the Institution of Electrical Engineers **113**, 1151 (1966).
2. M. L. Gorodetsky, A. A. Savchenkov, and V. S. Ilchenko, Opt. Lett. **21**, 453 (1996).
3. D. K. Armani, T. J. Kippenberg, S. M. Spillane, et al., Nature **421**, 925 (2003).
4. H. Lee, T. Chen, J. Li, et al., Nature Photon **6**, 369 (2012).
5. J. Geng, L. Yang, S. Zhao, et al., Opt. Express, OE **28**, 32907 (2020).
6. Y.-H. Lai, M.-G. Suh, Y.-K. Lu, et al., Nature Photonics **14**, 345 (2020).
7. C. Ropp, W. Zhu, A. Yulaev, et al., Light Sci Appl **12**, 83 (2023).
8. G. D. Martinez, C. Li, A. Staron, et al., Nat Commun **14**, 3501 (2023).
9. Y. Hu, M. Yu, B. Buscaino, et al., Nat. Photon. **16**, 679 (2022).
10. L. Chang, S. Liu, and J. E. Bowers, Nat. Photon. **16**, 95 (2022).
11. P. Trocha, M. Karpov, D. Ganin, et al., Science **359**, 887 (2018).
12. X. Zhang, K. Kwon, J. Henriksson, et al., Nature **603**, 253 (2022).
13. L.-M. Needham, C. Saavedra, J. K. Rasch, et al., Nature **629**, 1062 (2024).
14. J. Klein, M. Stern, G. Franchin, et al., 3D Printing and Additive Manufacturing **2**, 92 (2015).
15. F. Kotz, K. Arnold, W. Bauer, et al., Nature **544**, 337 (2017).
16. R. Dylla-Spears, T. D. Yee, K. Sasan, et al., Science Advances **6**, eabc7429 (2020).
17. X. Wen, B. Zhang, W. Wang, et al., Nat. Mater. **20**, 1506 (2021).
18. M. Li, L. Yue, A. C. Rajan, et al., Science Advances **9**, eadi2958 (2023).
19. P.-H. Huang, M. Laakso, P. Edinger, et al., Nat Commun **14**, 3305 (2023).
20. A. Legrain, T. G. Janson, J. W. Berenschot, et al., Journal of Applied Physics **115**, 214905 (2014).
21. W. S. Y. Wong, M. Li, D. R. Nisbet, et al., Science Advances **2**, e1600417 (2016).
22. H. Wang, H. Zhen, S. Li, et al., Science Advances **2**, e1600027 (2016).
23. S. Xu, Z. Yan, K.-I. Jang, et al., Science **347**, 154 (2015).
24. T. C. Shyu, P. F. Damasceno, P. M. Dodd, et al., Nature Mater **14**, 785 (2015).
25. J. Kim, J. A. Hanna, M. Byun, et al., Science **335**, 1201 (2012).
26. J.-H. Na, A. A. Evans, J. Bae, et al., Advanced Materials **27**, 79 (2015).
27. K. Chalapat, N. Chekurov, H. Jiang, et al., Adv Mater **25**, 91 (2013).
28. Y. Mao, Y. Zheng, C. Li, et al., Advanced Materials **29**, 1606482 (2017).
29. N. Lazarus and G. L. Smith, Advanced Materials Technologies **2**, 1700109 (2017).
30. D. Wu, Q. Zhang, G. Ma, et al., Optics and Lasers in Engineering **48**, 405 (2010).
31. E. Gärtner, J. Frühauf, U. Löschner, et al., Microsystem Technologies **7**, 23 (2001).
32. L. Michaeli, R. Gao, M. D. Kelzenberg, et al., Nat. Photon. **19**, 369 (2025).
33. G. Finkelstein-Zuta, Z. A. Arnon, T. Vijayakanth, et al., Nature **630**, 368 (2024).
34. L. M. Fonseca, G. V. Rodrigues, and M. A. Savi, International Journal of Mechanical Sciences **223**, 107316 (2022).
35. Y. Shimoda, K. Suto, S. Hayashi, et al., Also of interest 303 (2023).
36. H. E. Junfeng, W. E. N. Guilin, L. I. U. Jie, et al., International Journal of Mechanical Sciences **261**, 108690 (2024).
37. Y. Tao, J. M. Boss, B. A. Moores, et al., Nat Commun **5**, 3638 (2014).
38. I. Pikovski, M. R. Vanner, M. Aspelmeyer, et al., Nature Phys **8**, 393 (2012).
39. H. Yang, H. Miao, D.-S. Lee, et al., Phys. Rev. Lett. **110**, 170401 (2013).
40. G. Guccione, M. Hosseini, S. Adlong, et al., Phys. Rev. Lett. **111**, 183001 (2013).
41. L. Yang, T. Carmon, B. Min, et al., Applied Physics Letters **86**, Art (2005).
42. Y. Zhu, A. Ghrayeb, J. Yu, et al., Small **20**, 2400059 (2024).
43. N. Jin, C. A. McLemore, D. Mason, et al., Optica, OPTICA **9**, 965 (2022).
44. T. Plaskocinski, Y. Arita, G. D. Bruce, et al., Applied Physics Letters **123**, 081106 (2023).
45. J. C. Knight, G. Cheung, F. Jacques, et al., Optics Letters **22**, 1129 (1997).